\documentclass[prb,
                    preprint
                    ]{revtex4}

\usepackage{epsfig}
\usepackage{amsmath}
\usepackage{setspace}
\usepackage{subfigure}
\usepackage{color}

\def\e{\begin{equation}}
\def\f{\end{equation}}

\begin{document}


\title{Experimental verification of broadband cloaking using a volumetric cloak composed of periodically stacked cylindrical transmission-line networks}

\author{Pekka~Alitalo$^{1,2}$, Fr\'ed\'eric Bongard$^{2}$, Jean-Francois Z\"urcher$^{2}$, Juan Mosig$^{2}$, Sergei~Tretyakov$^{1}$}

\affiliation{$^1$~Department of Radio Science and Engineering /
SMARAD Center of Excellence\\ TKK Helsinki University of
Technology\\ P.O. Box 3000, FI-02015 TKK, Finland\\
$^2$~Laboratory of Electromagnetics and Acoustics (LEMA), Ecole
Polytechnique F\'ed\'erale de Lausanne (EPFL)
B\^atiment ELB, Station 11, CH-1015 Lausanne, Switzerland\\
}

\maketitle


\begin{center}
\section*{Abstract}
\end{center}

Cloaking using a volumetric structure composed of stacked
two-dimensional transmission-line networks is verified with
measurements. The measurements are done in a waveguide, in which
an array of metallic cylinders is inserted causing a short-circuit
in the waveguide. The metal cylinders are cloaked using a
previously designed and simulated cloak that ``hides'' the
cylinders and thus enables wave propagation inside the waveguide.



\newpage


The reduction of an object's total scattering cross section (SCS)
from electromagnetic waves impinging on the object from arbitrary
directions, often referred to as cloaking, has been the subject of
many works in the recent literature after the publication of some
seminal papers.\cite{Alu,Leonhardt,Pendry} The number of
scientific papers devoted to the study of this phenomenon is
already huge and therefore we will not review all the various
cloaking techniques here, but instead, we suggest the reader to
peruse the recent review paper by Al$\rm{\grave{u}}$ and
Engheta,\cite{Alu_review} and the references therein. The cloaking
phenomenon that is studied in this paper is achieved with the use
of so-called transmission-line networks. This approach to cloaking
has been studied analytically, numerically and experimentally in
some of our recent
papers.\cite{Alitalo_cloak_TAP,Alitalo_cloak_iWAT08,Alitalo_cloak_URSI08,Alitalo_cloak_META08,Alitalo_cloak_META}

In this paper, we experimentally demonstrate the cloaking
phenomenon using a cylindrically shaped network of transmission
lines. The ``cloak'' is a volumetric structure, composed of
several two-dimensional networks that are stacked on top of each
other. The cloak design that is used here was presented and
studied numerically recently.\cite{Alitalo_cloak_META08} A
waveguide environment has been chosen here for the measurements
since it allows to address the broadband behavior of the
considered cloak through the measurements of reflection and
transmission coefficients as functions of the frequency.

\begin{figure} [h!]
\centering \subfigure[]{\epsfig{file=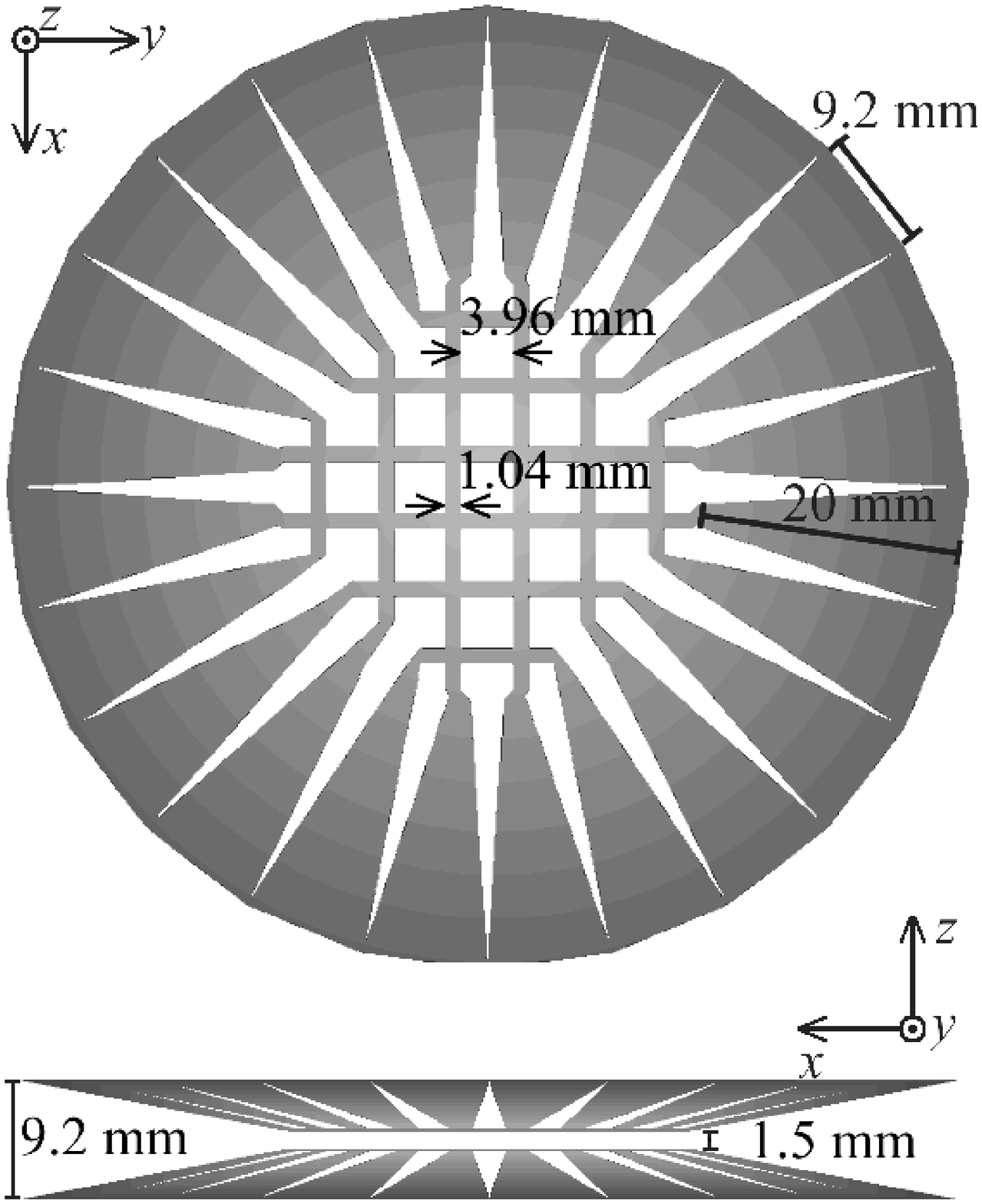,
width=0.325\textwidth}} \subfigure[]{\epsfig{file=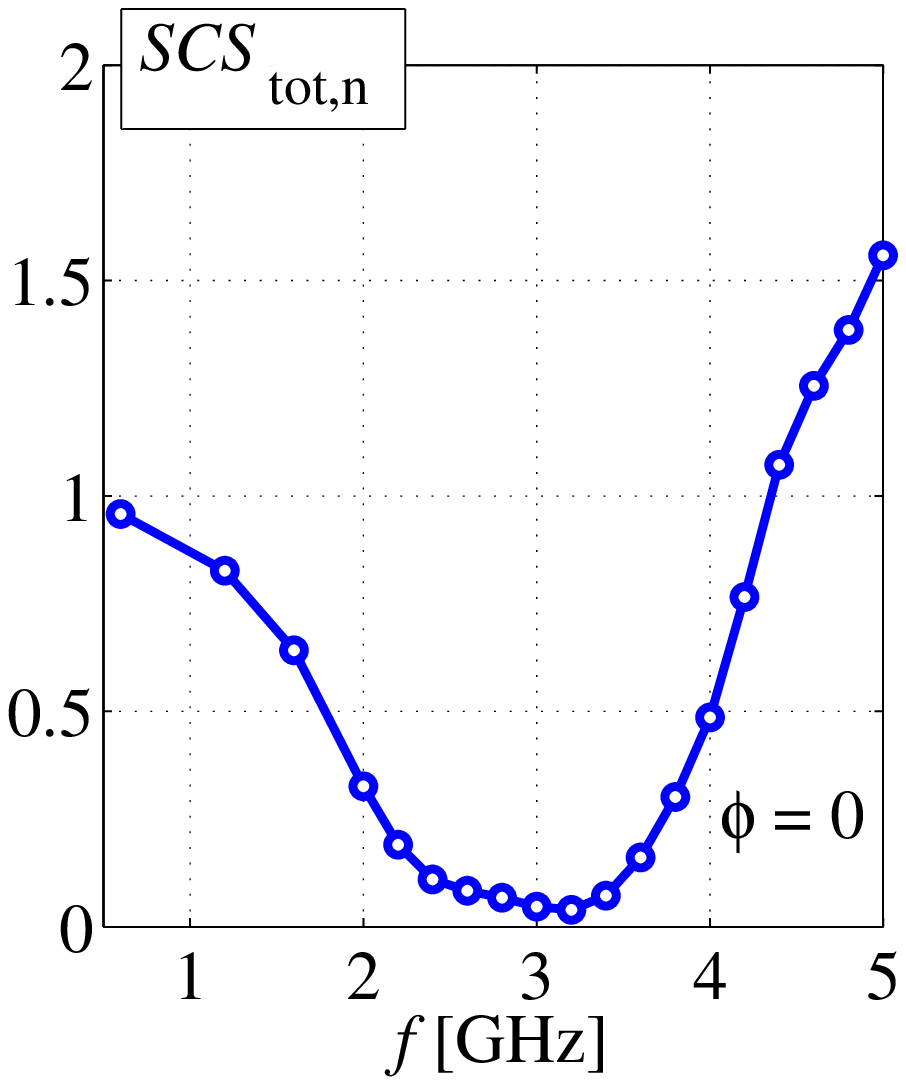,
width=0.25\textwidth}} \hspace{-0.55cm}
\subfigure[]{\epsfig{file=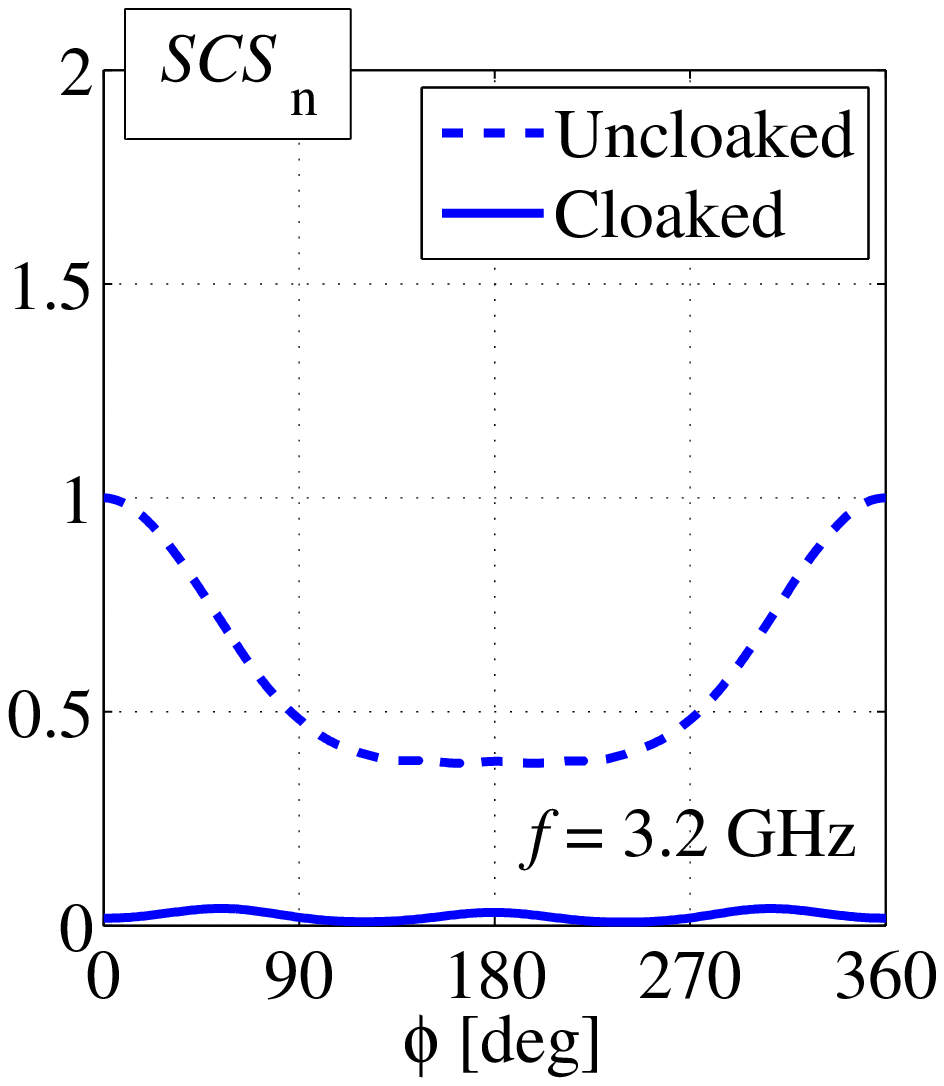,
width=0.25\textwidth}}\caption{(a)~Cloak structure and dimensions
in the $xy$-plane (top) and in the $xz$-plane (bottom).
(b)~Full-wave simulated \textit{total} scattering cross section of
the cloaked object, normalized to that of the uncloaked object.
(c)~Full-wave simulated scattering cross sections of uncloaked and
cloaked objects at the frequency of 3.2~GHz. $\phi$ is the angle
in the $xy$-plane and the plane wave illuminating the cloak
travels to the $+x$-direction, i.e., in the direction $\phi=0$.}
\label{cloak}
\end{figure}

The cloak that is used here is periodic in the vertical direction
($z$-direction), with the dimensions of a single period as
illustrated in Fig.~\ref{cloak}a. The period $d$ of the network is
5~mm and the width and height of the transmission lines are shown
in the figure. These dimensions have been found suitable for
optimal cloak operation around the frequency of 3~GHz, by
conducting full-wave simulations.\cite{Alitalo_cloak_META08} The
cloak is designed to work for TE-polarized waves with electric
field $E$ parallel to the $z$-axis. The cloak operation has been
previously studied with full-wave
simulations,\cite{Alitalo_cloak_META08} demonstrating the large
bandwidth where the total scattering cross section of a
two-dimensional array of metallic cylinders (the reference object
that we want to cloak) is reduced. In these simulations all the
metal objects such as the cloak itself and the cloaked object were
modelled as perfect electric conductor (PEC) in order to reduce
the simulation time. Cloaking with this type of device is not
limited to two-dimensional arrays but works similarly also for
three-dimensional objects, such as meshes. The only limitation is
that the cloaked object must fit inside the cloak structure, i.e.,
within the space outside the transmission
lines.\cite{Alitalo_cloak_TAP}


For the studied cloak structure, the cloaking phenomenon is first
confirmed by conducting full-wave simulations with Ansoft
HFSS\cite{Ansoft} of a cloaked and uncloaked object and
calculating the far-field scattering cross sections for both
cases.\cite{Alitalo_cloak_META08} In these simulations the cloak
is considered to be periodic and infinite along the $z$-direction.
See Fig.~\ref{cloak}b for the frequency dependence of the total
SCS of the cloaked object, normalized by the total SCS of the
uncloaked object (solid line). Fig.~\ref{cloak}c presents the
angular dependence of the scattering cross sections of cloaked and
uncloaked objects at the frequency of 3.2~GHz. The data presented
in Fig.~\ref{cloak}c is normalized to the maximum value of the
uncloaked object's SCS.

The cloak shown in Fig.~\ref{cloak}a was manufactured by etching
from 200~$\rm{\mu m}$ thick metal sheets composed of bronze
beryllium (BzBe). For stacking the cloak parts on top of each
other, we use layers of dielectric foam (Rohacell) with relative
permittivity $\varepsilon_{\rm r}\approx 1.05$.


The measurements are conducted with a modified aluminium WR-340
waveguide, with the inner dimensions 435~mm $\times$ 86.36~mm
$\times$ 36.8~mm (length $\times$ width $\times$ height). The
height is determined by the fact that it must be a multiple of the
cloak period in the vertical ($z$-) direction. In this case we use
4 networks on top of each other, i.e., 4 $\times$
9.2~mm~$=36.8$~mm. Transmission through the waveguide is measured
with standard coaxial probes with the distance from the waveguide
(front and back) walls being 23~mm. The height of the probes also
equals 23~mm. These values, with which good matching between the
probes and the waveguide are obtained, were found with full-wave
simulations of the empty waveguide.

\begin{figure} [h!]
\centering {\epsfig{file=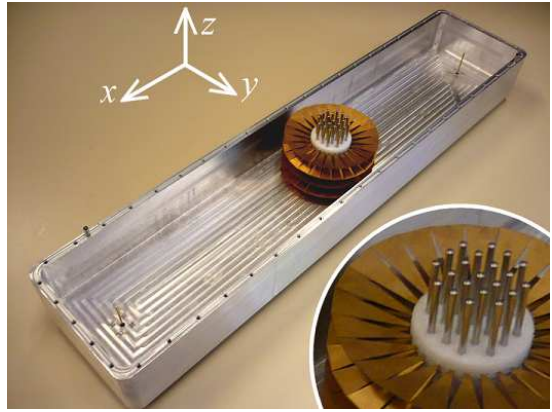, width=0.45\textwidth}}
\caption{Photograph of the waveguide with the cloaked object and
the cloak inside. The top wall of the waveguide is removed for
clarity. The inset shows a magnification of the cloaked object
enclosed by the cloak.} \label{Photo}
\end{figure}

The object that is supposed to be cloaked is a two-dimensional
array of metallic cylinders that fits inside the cloak structure.
The cylinders have a diameter of 2~mm and they are connected with
the bottom and top walls of the waveguide to create a
short-circuit inside the waveguide. See Fig.~\ref{Photo} for a
photograph of the waveguide with the cloak and the metal cylinders
inside.



\begin{figure} [h!]
\centering \subfigure[]{\epsfig{file=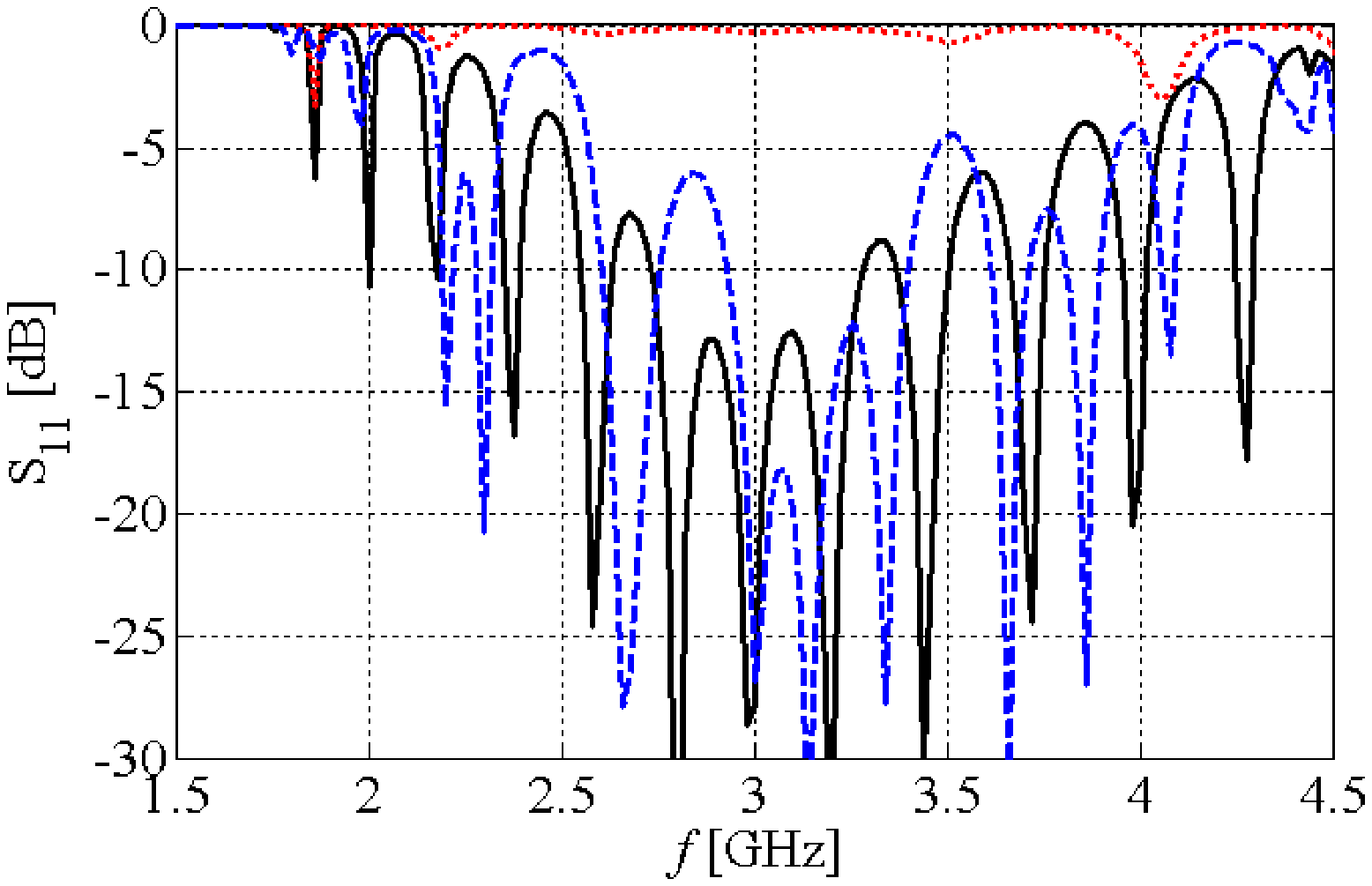,
width=0.45\textwidth}} \subfigure[]{\epsfig{file=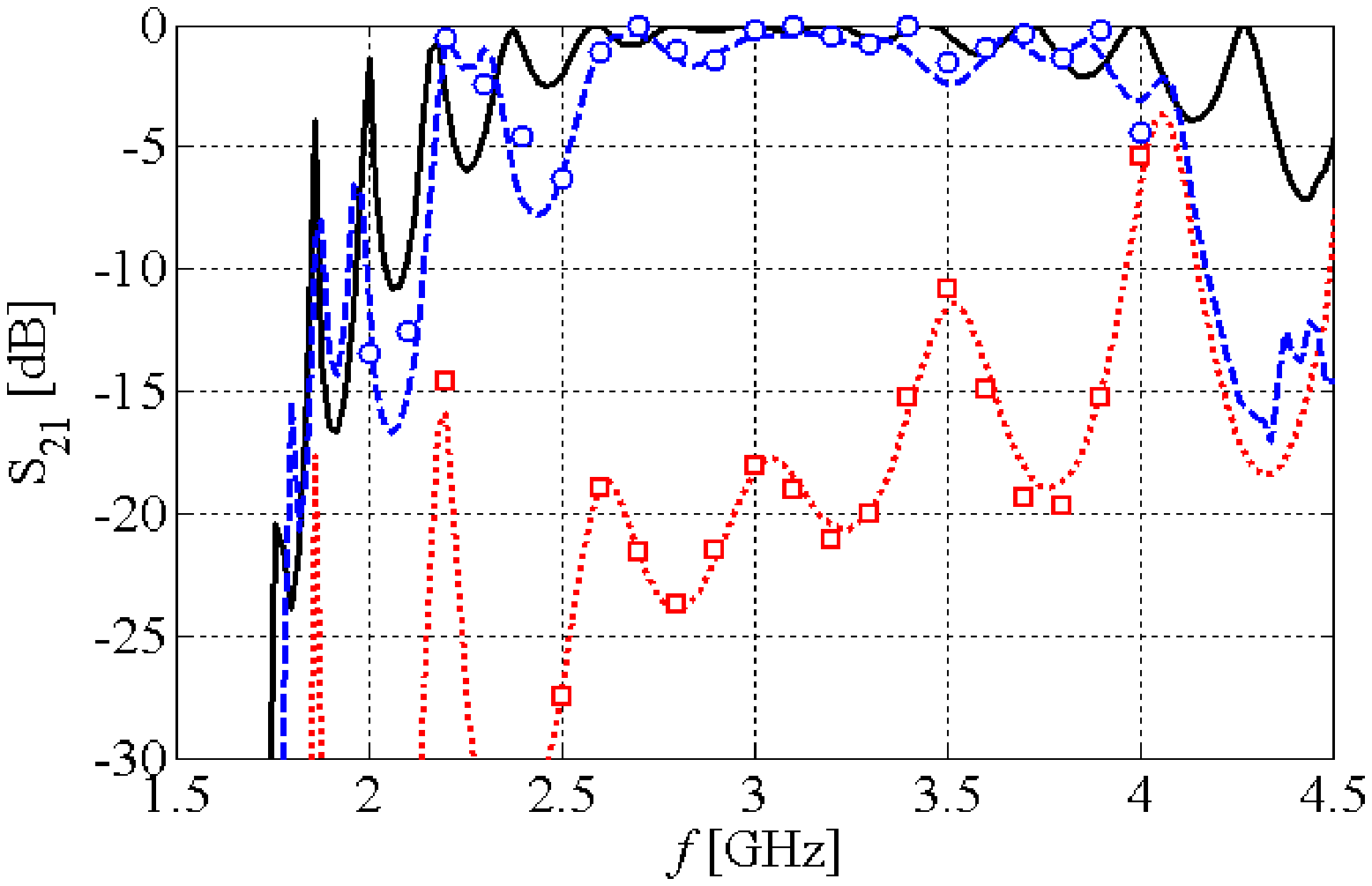,
width=0.45\textwidth}} \caption{Measured reflection (a) and
transmission (b) for the empty waveguide (solid line), waveguide
with the uncloaked object inside (dotted line) and waveguide with
the cloaked object inside (dashed line). For comparison the
full-wave simulated transmission corresponding to the uncloaked
(squares) and cloaked (circles) cases are also shown in (b).}
\label{S-param}
\end{figure}

The measured reflection and transmission for the empty waveguide
are shown in Fig.~\ref{S-param}, demonstrating that the empty
waveguide section has low reflections and high transmittance for
the frequencies of interest, i.e., around 3~GHz, where the cloak
is supposed to work the best. With the uncloaked object inside the
waveguide, the transmission is less than $-15$~dB around this
frequency. With the cloaked object, the measured transmission and
reflection agree well with those of the empty waveguide,
especially in the frequency band from 2.5~GHz to 4~GHz.


The measurement results are in good agreement with the previous
simulations of the cloak, experimentally confirming the broadband
cloaking phenomenon also for real structures. As the fundamental
mode inside the waveguide can be considered to be a sum of two
plane waves with the incidence angles varying as functions of the
frequency, these results also give further confirmation of the
isotropy of the cloak. For comparison, also full-wave simulation
results for cloaked and uncloaked objects inside the waveguide are
shown in Fig.~\ref{S-param}b.

The phase of the measured $S_{\rm 21}$ in the empty waveguide and
the cloaked case are obviously different. For example, in the
frequency range from 2.5~GHz to 3.3~GHz, the absolute value of the
phase difference between these two cases varies between 0 and 45
degree. The envelope of the phase difference curve grows in
magnitude with increasing frequency. This phase difference is due
to the fact that the wave number inside the cloak is slightly
different than in free space.\cite{Alitalo_cloak_TAP} As
previously discussed\cite{Alitalo_cloak_TAP} and also demonstrated
by Figs.~\ref{cloak}b and \ref{cloak}c, in many practical
situations this non-ideality does not prevent efficient cloaking.


We have presented a waveguide-based measurement procedure, with
which we can, in a convenient way, study the cloaking phenomenon
of certain types of cloaks. In this paper we have studied the
performance of a volumetric microwave cloak composed of layered
two-dimensional transmission-line networks. The broadband cloaking
phenomenon, which has been studied before with numerical
simulations, is now confirmed with measurements.

\section*{Acknowledgements}

This work has been partially funded by the Academy of Finland and
TEKES through the Center-of-Excellence program. During this work
P.~Alitalo was an invited researcher at EPFL-Switzerland.
P.~Alitalo acknowledges financial support by GETA, the Emil
Aaltonen Foundation, and the Nokia Foundation.

\clearpage

\end{document}